\documentclass{ieeeaccess}
\usepackage{cite}
\usepackage{amsmath,amssymb,amsfonts}
\usepackage{algorithmic}
\usepackage{graphicx}
\usepackage{textcomp}
\usepackage{booktabs} 
\usepackage{multirow} 
\usepackage{comment}
\usepackage{url}

\usepackage{bm}
\makeatletter
\AtBeginDocument{\DeclareMathVersion{bold}
\SetSymbolFont{operators}{bold}{T1}{times}{b}{n}
\SetSymbolFont{NewLetters}{bold}{T1}{times}{b}{it}
\SetMathAlphabet{\mathrm}{bold}{T1}{times}{b}{n}
\SetMathAlphabet{\mathit}{bold}{T1}{times}{b}{it}
\SetMathAlphabet{\mathbf}{bold}{T1}{times}{b}{n}
\SetMathAlphabet{\mathtt}{bold}{OT1}{pcr}{b}{n}
\SetSymbolFont{symbols}{bold}{OMS}{cmsy}{b}{n}
\renewcommand\boldmath{\@nomath\boldmath\mathversion{bold}}}
\makeatother

\def\BibTeX{{\rm B\kern-.05em{\sc i\kern-.025em b}\kern-.08em
    T\kern-.1667em\lower.7ex\hbox{E}\kern-.125emX}}

\begin{document}

\history{}
\doi{}

\title{Spectrotemporal Feature Extraction in EHG Signals and Tocograms for Enhanced Preterm Birth Prediction}

\author{\uppercase{Senith Jayakody}\authorrefmark{1},
\uppercase{Kalana Jayasooriya}\authorrefmark{1}, 
\uppercase{Sashini Liyanage}\authorrefmark{1},
\uppercase{Roshan Godaliyadda}\authorrefmark{1,2},
\uppercase{Parakrama Ekanayake}\authorrefmark{1,2},
\uppercase{Isuru Nawinne}\authorrefmark{3},
\uppercase{Chathura Rathnayake}\authorrefmark{4},
}

\address[1]{Multidisciplinary AI Research Center, University of Peradeniya, Peradeniya 20400, Sri Lanka}
\address[2]{Department of Electrical and Electronic Engineering, University of Peradeniya, Peradeniya 20400, Sri Lanka}
\address[3]{Department of Computer Engineering, University of Peradeniya, Peradeniya 20400, Sri Lanka}
\address[4]{Department of Obstetrics and Gynaecology, University of Peradeniya, Peradeniya 20400, Sri Lanka}
\tfootnote{This work was supported by the University Research Council (URC), University of Peradeniya, under Grant URC/2025/MRG/33 and URC/2022/MRG/207.}

\markboth
{S. Jayakody \headeretal: Spectrotemporal Feature Extraction in EHG Signals and Tocograms for Enhanced Preterm Birth Prediction}
{S. Jayakody \headeretal: Spectrotemporal Feature Extraction in EHG Signals and Tocograms for Enhanced Preterm Birth Prediction}

\corresp{Corresponding author: Senith Jayakody (e-mail: senith@eng.pdn.ac.lk).}

\begin{abstract}
Preterm birth (PTB), defined as delivery before 37 weeks of gestation, is a leading cause of neonatal mortality and long term health complications. Early detection is essential for enabling timely medical interventions. Electrohysterography (EHG) and tocography (TOCO) are promising non invasive tools for PTB prediction, but prior studies often suffer from class imbalance, improper oversampling, and reliance on features with limited physiological relevance. This work presents a machine learning (ML) pipeline incorporating robust preprocessing, physiologically grounded feature extraction, and rigorous evaluation. Features were extracted from EHG (and TOCO) signals using Mel frequency cepstral coefficients, statistical descriptors of wavelet coefficients, and peaks of the normalized power spectrum. Signal quality was enhanced via Karhunen Loève Transform (KLT) denoising through eigenvalue based subspace decomposition. Multiple classifiers, including Logistic Regression, Support Vector Machines, Random Forest, Gradient Boosting, Multilayer Perceptron, and CatBoost, were evaluated on the TPEHGT dataset. The CatBoost classifier with KLT denoising achieved the highest performance on fixed interval segments of the TPEHGT dataset, reaching 97.28\% accuracy and an AUC of 0.9988. Ablation studies confirmed the critical role of both KLT denoising and physiologically informed features. Comparative analysis showed that including TOCO signals did not substantially improve prediction over EHG alone, highlighting the sufficiency of EHG for PTB detection. These results demonstrate that combining denoising with domain-relevant features can yield highly accurate, robust, and clinically interpretable models, supporting the development of cost-effective and accessible PTB prediction tools, particularly in low-resource healthcare settings.
\end{abstract}

\begin{keywords}
Electrohysterography (EHG), Preterm birth detection, Uterine contractions, Biomedical signal analysis, Machine learning (ML)
\end{keywords}

\titlepgskip=-21pt

\maketitle

\section{Introduction}
\label{sec:introduction}
\PARstart{P}{reterm} birth (PTB), defined as delivery before 37 weeks of gestation \cite{whoPretermBirth, cdc_preterm_def}, remains a significant global health challenge. Preterm labour (PTL) refers to the onset of uterine contractions with cervical changes before 37 weeks, while PTB is the final delivery event that follows. PTL is driven by processes such as infection, inflammation, uterine overdistension, and hormonal activation, leading to cervical ripening and increased uterine contractility, which may ultimately result in PTB \cite{PTL1, ptl_2}. In 2020 alone, approximately 13.4 million babies, roughly one in ten live births, were born prematurely \cite{whoPretermBirth}. Prematurity is the leading cause of death among children under five, highlighting its critical impact on early childhood survival \cite{whoPretermBirth, Ohuma2023}. Even among survivors, PTB is associated with severe health risks, including developmental delays, disabilities, and an increased likelihood of chronic diseases such as diabetes and heart conditions later in life \cite{whoPretermBirth}. PTB imposes significant psychological and physiological stress on mothers and also places a considerable burden on healthcare systems due to the need for prolonged neonatal care and resource intensive interventions \cite{Lasiuk2013}. Addressing this issue requires urgent improvements in both the care provided to preterm infants and preventive measures, particularly in maternal and child healthcare, to enhance survival outcomes.

Cost effective interventions have the potential to prevent up to 75\% of preterm related deaths, yet access to appropriate care remains highly unequal based on country income \cite{whoPretermBirth}. In low income regions, nearly half of the infants born at or before 32 weeks die due to inadequate medical resources, whereas in high income settings, survival rates are significantly higher. The disparity is even more pronounced for extremely preterm infants (born before 28 weeks), where more than 90\% die within days in low income countries, compared to less than 10\% in high income regions \cite{whoPretermBirth}. Early detection of PTB is crucial for improving survival rates and reducing complications, as it enables better resource allocation and timely medical interventions for both the mother and the baby.

There has been growing interest in research focused on detecting PTBs. Many studies aim to develop cost effective and accessible solutions, especially for low and middle income regions where most cases occur. One promising approach involves using non invasive electrohysterography (EHG) recordings to detect PTL \cite{XU2022103231, GarciaCasado2018}. EHG signals capture the electrical activity of the uterus through the abdominal surface, providing insight into the frequency, duration, and intensity of uterine contractions, which are key indicators of labor progression \cite{R2020}. Alongside EHG, tocography (TOCO) is also commonly used. TOCO tracks the mechanical pressure on the abdominal wall, helping to assess the strength and timing of contractions \cite{Chen2024}.

However, despite ongoing research, the field faces a major bottleneck: the lack of large, diverse, and representative datasets, with particularly limited availability in regions such as South Asia, and a general absence of historical medical and socioeconomic information. Some of the past studies have used private or publicly inaccessible datasets \cite{GarciaCasado2018, Mischi2018}, making replication and benchmarking difficult. Even the most widely used public datasets, such as the TPEHGT \cite{2018Dataset} dataset, which contains thirteen preterm and thirteen term records, and the TPEHG \cite{2012Dataset} dataset, which includes 38 preterm and 262 term records, were collected exclusively in Slovenia. As a result, they not only suffer from small sample sizes or significant class imbalance but also introduce an inevitable regional and geographic bias, limiting the generalizability of findings to populations in other parts of the world. Due to the low incidence of PTB, most studies have had to work with limited data.

To overcome data imbalance, many researchers apply different augmentation techniques \cite{XU2022103231}. However, as Vandewiele et al. \cite{VANDEWIELE2021101987} pointed out, most high performing benchmark models applied oversampling~\cite{HeGarcia2009} before splitting the dataset into training and testing sets. This approach inflates performance metrics, making the results unreliable. Vandewiele et al. highlighted this by revisiting the original study \cite{2018Dataset} behind the widely used public TPEHGT dataset. The study reported an AUC of 99.44 using quadratic discriminant analysis (QDA) with adaptive synthetic (ADASYN) oversampling, but when oversampling was correctly applied after data splitting, performance dropped to 62.36. Their findings exposed major flaws in past studies that adopted these evaluation practices, constituting the majority of the work, emphasizing the need for more rigorous and realistic modeling in this critical domain.

Due to the small sample sizes in existing public datasets, most predictions have relied on relatively simple, classical machine learning (ML) methods. Although recent advances have introduced deep learning (DL) approaches, only a few models have been explored so far, mainly due to limited data availability \cite{Zhang2024}. Moreover, DL models often lack explanability and transparency, providing limited insight into underlying physiological mechanisms, which can hinder their adoption in clinical settings \cite{Jager2024}. As a result, traditional ML methods with manual feature extraction remain the common approach. 

However, overly inflated results caused by improper oversampling have led many traditional ML studies to prioritize large sets of extracted features over contextual feature selection, producing artificially inflated accuracies that offer little real world applicability. As a result, PTB prediction remains an open and largely unsolved problem, as most past work has reported inflated results that are not truly applicable to the real problem. Hence, as noted by Vandewiele et al., a re-examination from the basics is warranted. 


From a physiological perspective, specific frequency bands in EHG signals are known to reflect meaningful uterine and maternal activity. The 0.08–1.0 Hz range is primarily linked to uterine physiological activity, while the 1.0–2.2 Hz, 2.2–3.5 Hz, and 3.5–5.0 Hz bands are associated with the maternal heart’s influence on the uterus, including the heart rate and its second and third harmonics \cite{Jager2024}. Notably, the 1.0–2.2 Hz band, which reflects maternal heart induced electrical and mechanical uterine activity, tends to show higher amplitude during term pregnancies when labor is still distant, and lower amplitude as delivery approaches. Interestingly, this band is also observed to have low amplitudes during preterm pregnancies, even when delivery is not expected soon \cite{2018Dataset}. This behavior suggests that specific sub bands in the frequency domain carry predictive information, emphasizing the need for targeted feature extraction to capture the underlying physiological signals linked to PTB.

In this paper, we revisit existing publicly available datasets and address key gaps related to improper oversampling and the limited focus on extracting physiologically meaningful frequency domain features from EHG signals. Our approach introduces a reliable and reproducible pipeline for PTB detection, evaluated on a second publicly available dataset to assess consistency. The main contributions of this work are:
\begin{itemize}
    \item Noise subspace separation using the Karhunen Loève Transform (KLT) with eigenvalue decomposition to enhance EHG signal quality. KLT adapts filtering based on the measured signal's noise characteristics, avoiding reliance on fixed band assumptions and improving practical applicability.
    \item Extraction of physiologically relevant features inspired by acoustic signal processing, including Mel Frequency Cepstral Coefficients (MFCCs), wavelet based spectral descriptors, and Peak Amplitudes (PA) of the normalized power spectrum, capturing both uterine activity and spectral characteristics relevant to PTB.
    \item Design of the feature extraction and evaluation pipeline to support interpretability, enabling physiologically meaningful insights and explainable model predictions.
    \item Systematic evaluation of EHG with and without TOCO signals, demonstrating that EHG alone is sufficient for robust PTB prediction, which is particularly relevant for low resource clinical settings.
    \item Evaluation of the proposed features across two public datasets (TPEHG and TPEHGT), demonstrating consistent performance improvements over state of the art models.
\end{itemize}

To present our study clearly, the remainder of this paper is organized as follows. Section \ref{sec:method} describes the datasets, preprocessing steps, feature extraction, and classification models used in the proposed approach. Section \ref{sec:results} presents the experimental results and discussion, compares them with existing methods, and includes an ablation study. Finally, Section \ref{conclusion} summarizes the main contributions of this work.

\section{Materials and Methods}
\label{sec:method}
\begin{figure*}[tb]
    \centering
    \includegraphics[width=1\linewidth]{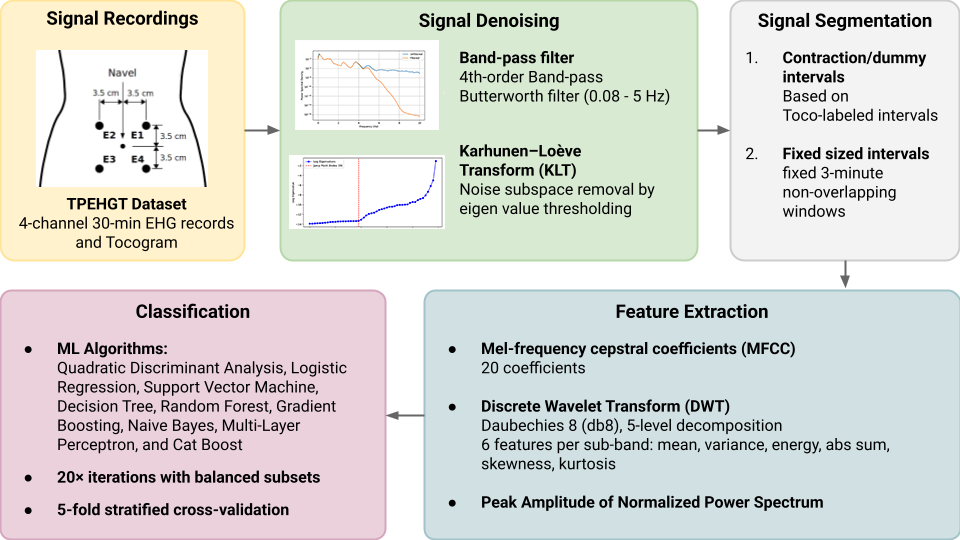}
    \caption{Proposed method for classifying PTBs using EHG recordings}
    \label{fig:method}
\end{figure*}

An overview of the proposed method is shown in Fig.~\ref{fig:method}, with details discussed in the following sections.

\subsection{Dataset}
This study employs two publicly available EHG datasets hosted on PhysioNet: the TPEHGT dataset, published in 2018 \cite{2018Dataset}, and the TPEHG dataset, published in 2012 \cite{2012Dataset}. Both datasets were developed at the Faculty of Computer and Information Science, University of Ljubljana, Slovenia, with recordings collected at the Department of Obstetrics and Gynecology, University Medical Center Ljubljana. The main experiments and results presented in the following sections were conducted using the widely adopted TPEHGT dataset. To further evaluate the proposed pipeline on an additional dataset, we report results on the TPEHG dataset.

The TPEHGT dataset contains 31 uterine recordings: thirteen from preterm pregnancies, thirteen from term pregnancies, and five from non pregnant women. EHG signals were recorded using four electrodes placed in two horizontal rows symmetrically above and below the navel, spaced 7 cm apart. Each 30-minute recording includes four channels: three EHG signals (differential electrode pairs) and one tocogram signal. The preterm subset has a mean delivery time of 33.7 weeks and provides 47 annotated contraction intervals and 47 dummy intervals. The term subset has a mean delivery time of 38.1 weeks with 53 annotated contraction intervals and 53 dummy intervals.

The TPEHG dataset contains 300 uterine EHG recordings, each lasting 30 minutes and comprising three channels. It includes 38 preterm and 262 term pregnancies, with 162 recordings obtained before the 26th week of gestation and the remainder collected at later stages.

\subsection{Preprocessing}
\subsubsection{Band-pass filtering}
The primary dataset used for model development is the TPEHGT dataset \cite{2018Dataset}, which includes eight signals per recording: three unfiltered EHG channels, one unfiltered TOCO signal, and their corresponding filtered versions. The filtering was performed using a fourth order band pass Butterworth filter with a passband of 0.08–5.0 Hz. For feature extraction, we used the filtered EHG channels, as they retain physiologically relevant frequency components while attenuating high frequency noise and low frequency drift. 

For validation, we used the TPEHG dataset \cite{2012Dataset}, which contains uterine EHG recordings from 300 pregnancies. To ensure consistency with the TPEHGT dataset, we applied a fourth order Butterworth filter with a 0.08-5.0 Hz range to the unfiltered EHG channels in the TPEHG dataset, matching the preprocessing approach used for TPEHGT. 

As discussed earlier, the EHG signal contains distinct physiological components: the 0.08–1.0 Hz band is mainly linked to uterine activity, while the 1.0–5.0 Hz range captures maternal cardiac influence, including the heart rate and its harmonics \cite{Jager2024}. While these bands offer physiological insights, we chose not to limit feature extraction to predefined sub bands. Instead of hand picking frequency bands and assuming their relevance, we retained the entire physiologically relevant range (0.08–5 Hz) for feature extraction. This approach allows the model to learn the most predictive patterns directly from the data, without bias introduced by prior assumptions about which sub bands are informative. By using the full band, the KLT can optimally separate noise and overlapping components, enhancing signal quality and maximizing the discriminative information available for downstream classification.

\begin{figure}[tb]
    \centering
    \includegraphics[width=1\linewidth]{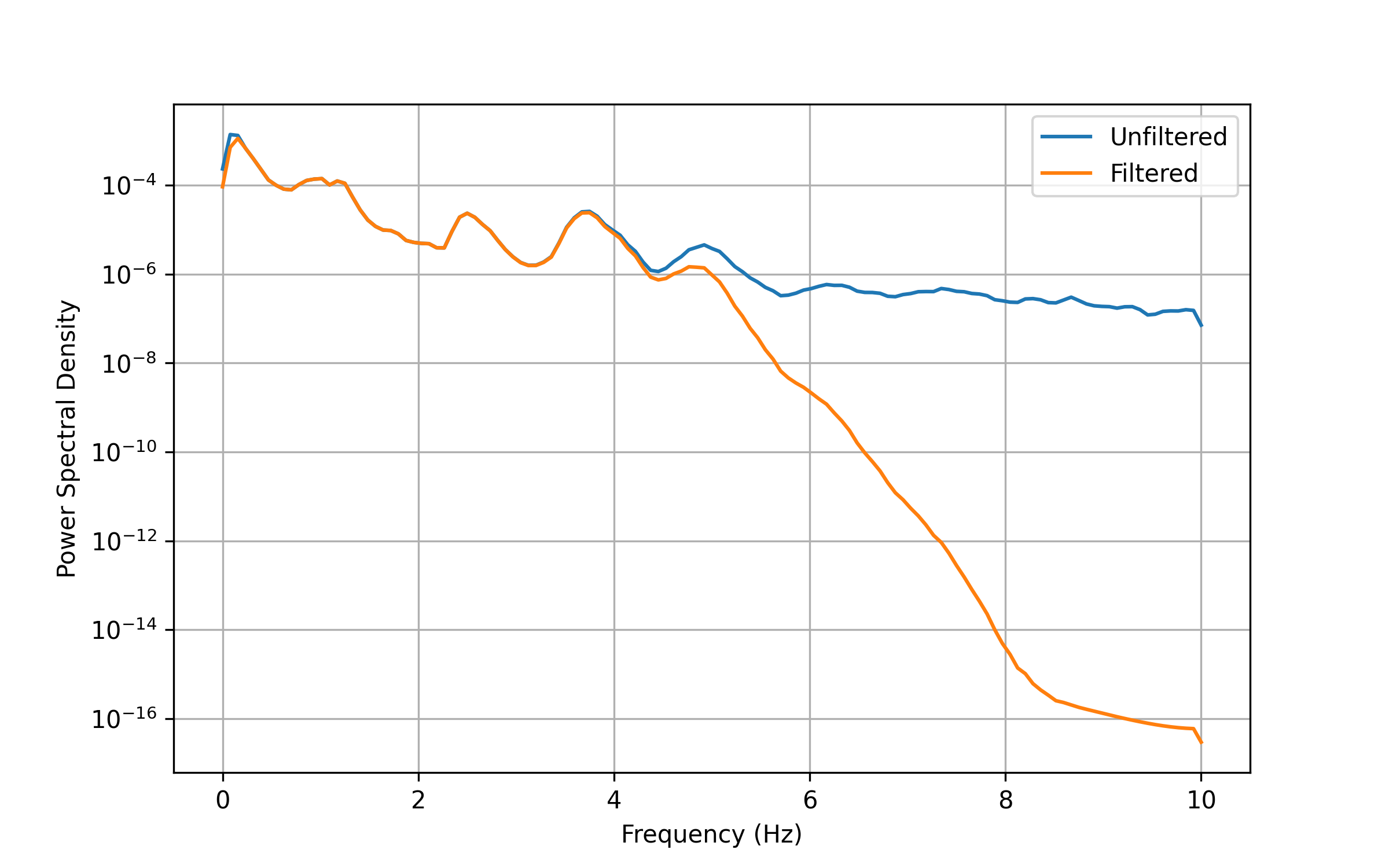}
    \caption{Example power spectral density (PSD) of the EHG signal before (blue) and after (orange) applying a 4th order Butterworth band pass filter (0.08–5.0 Hz)}
    \label{fig:filtered}
\end{figure}

To do this, we applied a coarse Butterworth bandpass filter in the 0.08–5 Hz range to remove outright noise. Fig.~\ref{fig:filtered} illustrates the PSD of the signal before and after filtering. The filtered signal was then used for spectral feature engineering. Subsequently, finer signal subspace separation and feature selection were performed using KLT as mentioned in Section \ref{subsec: KLT}. This hierarchical approach avoids biasing the system toward predefined bands, allowing the model to learn the most predictive frequency domain characteristics and justifying the use of KLT as noted in our contributions.

\subsubsection{Karhunen Loève Transform}
\label{subsec: KLT}

While band pass filtering suppresses out of band interference and isolates physiologically relevant frequency components, residual in band noise and artifacts may still remain and can degrade feature quality. To further refine the EHG signal, we apply a KLT based subspace projection to attenuate noise dominated components while preserving the structured content of the signal.



The KLT represents a signal in an orthogonal basis obtained from the eigendecomposition of its second order statistics. For a wide sense stationary process, the autocorrelation matrix is approximately Toeplitz, and its eigenvectors capture dominant correlation patterns. In contrast, white noise is approximately uncorrelated, leading to a covariance close to $\sigma^2 I$ with a relatively flat eigenvalue spectrum (i.e., no dominant directions). Motivated by this behavior, we separate a signal dominated subspace associated with larger eigenvalues from a noise dominated subspace associated with small, weakly varying eigenvalues.




Let $x[n]$ denote an input discrete-time signal of length $N$. Rather than forming the full $N\times N$ correlation matrix, we restrict the autocorrelation to a fixed lag $L=50$ to focus on local temporal structure and reduce computational complexity. The unbiased autocorrelation sequence is estimated as
\begin{equation}
r[k] = \frac{1}{N-k}\sum_{n=0}^{N-k-1} x[n]\,x[n+k], \quad k=0,1,\dots,L-1.
\end{equation}
Using the first $L$ lags, the autocorrelation matrix $R \in \mathbb{R}^{L \times L}$ is constructed as a symmetric Toeplitz matrix:
\begin{equation}
R =
\begin{bmatrix}
r[0] & r[1] & r[2] & \cdots & r[L-1] \\
r[1] & r[0] & r[1] & \cdots & r[L-2] \\
r[2] & r[1] & r[0] & \cdots & r[L-3] \\
\vdots & \vdots & \vdots & \ddots & \vdots \\
r[L-1] & r[L-2] & r[L-3] & \cdots & r[0]
\end{bmatrix}.
\end{equation}



The KLT basis is obtained from the eigendecomposition of $R$:
\begin{equation}
R = Q \Lambda Q^T,
\label{eq:klt_decomposition}
\end{equation}
where $\Lambda=\mathrm{diag}(\lambda_1,\lambda_2,\dots,\lambda_L)$ contains eigenvalues and $Q=[q_1,q_2,\dots,q_L]$ contains the corresponding eigenvectors. Eigenvalues are sorted in ascending order ($\lambda_1 \le \lambda_2 \le \cdots \le \lambda_L$). In implementation, eigenvectors are $\ell_2$ normalized to unit norm to form a well defined orthogonal projection.

\begin{figure}[tb]
    \centering
    \includegraphics[width=1\linewidth]{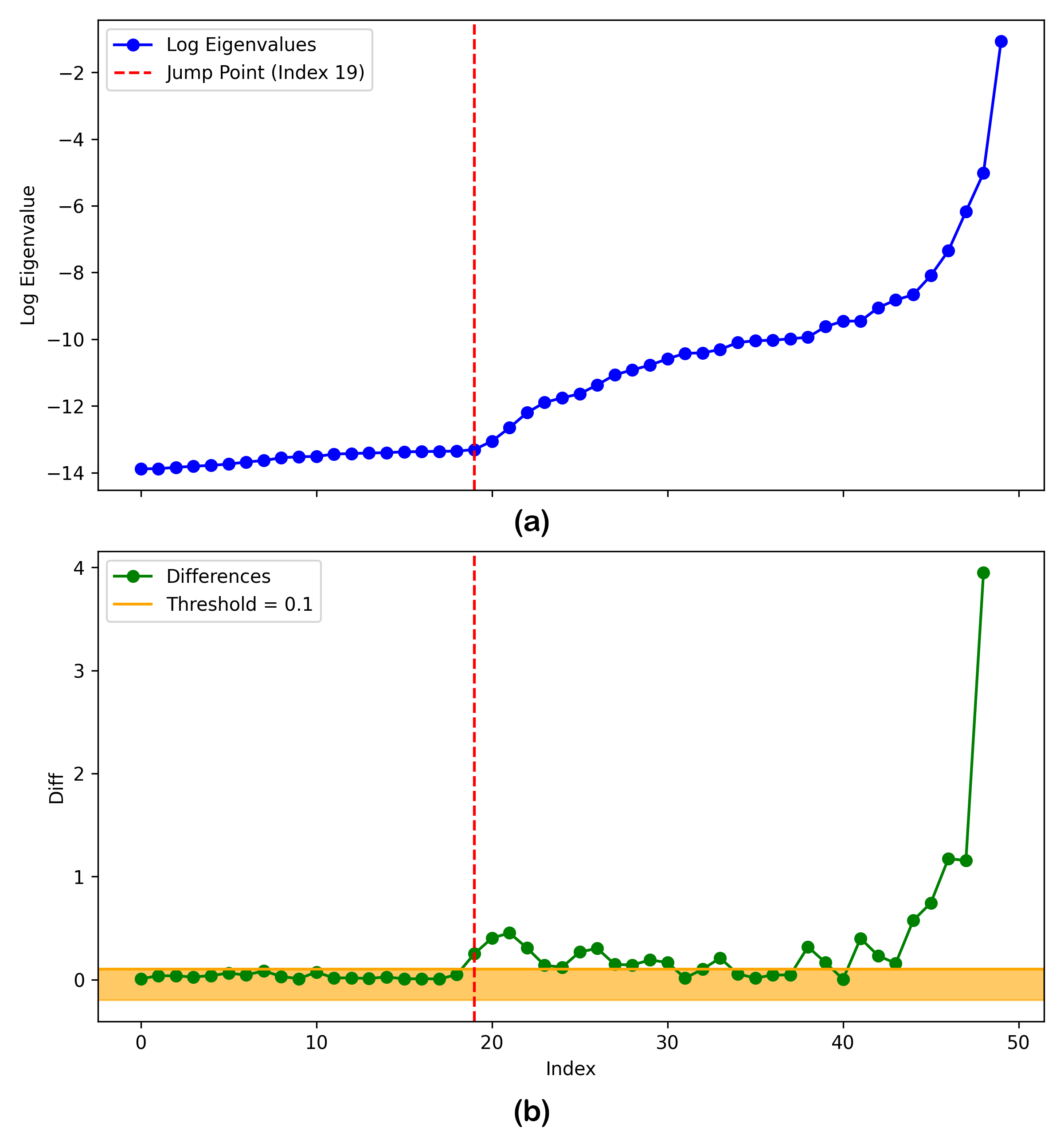}
    \caption{(a) Log-eigenvalue spectrum $\log(\lambda_i)$ in ascending order. (b) Consecutive log-eigenvalue differences $\Delta_i$. The 10\% jump threshold ($\tau=0.1$) identifies the transition from noise-dominated components (left) to signal-dominated components (right) retained for reconstruction.}
    \label{fig:eigenval_point}
\end{figure}

\textit{Subspace selection via a 10\% eigen jump:}
To identify the transition between noise dominated and signal dominated components, we analyze the differences between consecutive log eigenvalues:
\begin{equation}
\Delta_i = \log(\tilde{\lambda}_{i+1}) - \log(\tilde{\lambda}_i), \quad i=1,2,\dots,L-1,
\end{equation}
where $\tilde{\lambda}_i = |\lambda_i| + \epsilon$ (with a small $\epsilon$) is used for numerical stability. We define the transition index $k$ as the first instance where the log eigenvalue jump exceeds a 10\% criterion,
\begin{equation}
\Delta_k > \tau, \quad \tau = 0.1,
\end{equation}
which corresponds to approximately a 10\% increase in eigenvalue magnitude (since $e^{0.1}\approx 1.105$). If no such index exists, $k$ is set to the index of the maximum jump. Fig.~\ref{fig:eigenval_point} illustrates the log eigenvalue spectrum and the corresponding inter-eigenvalue differences. The eigenvectors associated with the larger eigenvalues beyond this transition are retained as the signal subspace:
\begin{equation}
Q_s = [q_{k+1}, q_{k+2}, \dots, q_L].
\end{equation}

\textit{Block wise denoising via subspace projection:}
The denoised signal is obtained by projecting the input onto the selected subspace. Specifically, the signal is partitioned into non overlapping blocks of length $L$, forming vectors $\mathbf{x}_m \in \mathbb{R}^{L}$. Each block is denoised using the orthogonal projector $Q_s Q_s^T$:
\begin{equation}
\widehat{\mathbf{x}}_m = Q_s Q_s^T \mathbf{x}_m.
\end{equation}
The final denoised signal $\widehat{x}[n]$ is obtained by concatenating the reconstructed blocks. This procedure suppresses contributions from the noise dominated subspace while retaining the dominant correlated structure of the EHG signal. The resulting noise filtered signal is then used for feature extraction and subsequent classification.

\subsection{Segmentation and Sample Construction}
\label{sec:segmentation}

This work evaluates two segmentation strategies that define the sample unit used for feature extraction and classification: (i) contraction-dummy interval segmentation provided by the TPEHGT dataset annotations, and (ii) fixed length blind segmentation into non overlapping 3-minute windows. In both cases, each segment is treated as one sample and inherits the record level label (term vs. preterm) of its originating pregnancy record.

\subsubsection{Contraction-dummy intervals}
\label{subsec:seg_contraction_dummy}
For the 2018 TPEHGT dataset, we use the contraction interval annotations provided with the dataset, together with the corresponding dummy (non contraction) intervals. Each annotated contraction interval and each dummy interval is treated as a separate sample, and the sample label is inherited from the mother/record label (term or preterm). Using the preterm subset yields $47$ contraction + $47$ dummy intervals ($94$ samples), while the term subset yields $53$ contraction + $53$ dummy intervals ($106$ samples). To address class imbalance for model training, term samples are randomly undersampled to match the $94$ preterm samples per iteration (details in Section~\ref{sec:evaluation}).

\subsubsection{Fixed 3-minute non overlapping intervals}
\label{subsec:seg_fixed3min}
To simulate a blind segmentation scenario without relying on contraction boundaries, each 30-minute record is partitioned into $10$ non overlapping windows of 3 minutes. We use 3-min windows because typical contractions last~1–2 min, and 3 min captures a full event plus baseline context. Each 3-minute window is treated as one sample and inherits the record label.
For TPEHGT (excluding non pregnant records), this produces $26 \times 10 = 260$ samples (130 term + 130 preterm).
For TPEHG (300 records: 38 preterm, 262 term), this produces $300 \times 10 = 3000$ windows in total (380 preterm + 2620 term). For balanced training iterations, all 380 preterm windows are used and 380 term windows are randomly selected per iteration.

\subsection{Feature Extraction}

\subsubsection{Peak amplitude of the normalized power spectrum} 


The PA of the normalized power spectrum characterizes the dominant spectral component within a specified frequency band. This feature is designed to capture relative spectral concentration while remaining invariant to absolute signal amplitude.

Electrical and mechanical activity induced by the maternal heart in the uterus tends to show higher amplitude during term pregnancies when labor is still distant. This amplitude typically decreases as the delivery approaches. Interestingly, a similar low amplitude is observed in preterm pregnancies, even when delivery is expected soon~\cite{2018Dataset}. This characteristic makes the PA a crucial predictor of preterm birth.


For a discrete time signal $x[n]$, the power spectral density (PSD) $P(f_i)$ is estimated using Welch’s method~\cite{Welch1967} over discrete frequency bins $f_i$. A frequency band $[f_{\text{low}}, f_{\text{high}}]$ is then selected, and the PSD values within this band are normalized by the total band power to ensure comparability across recordings of varying signal energy. The PA is defined as the maximum value of the band normalized PSD.
\begin{equation}
A_{\text{peak}} =
\max_{f_i \in [f_{\text{low}}, f_{\text{high}}]}
\frac{P(f_i)}{\sum_{f_j \in [f_{\text{low}}, f_{\text{high}}]} P(f_j)} 
\end{equation}

\subsubsection{Mel coefficients extraction}
Mel Frequency Cepstral Coefficients (MFCCs)~\cite{DavisMermelstein1980,Mel-low-frequency} represent the short term power spectrum of a signal and are commonly used for feature extraction based on the Short-Time Fourier Transform (STFT). Initially developed for speech recognition, MFCCs have gained attention in biomedical signal processing due to their ability to capture perceptually relevant spectral features. MFCCs employ a Mel scale filter bank, which maps the frequency axis nonlinearly, providing higher resolution for lower frequencies. This emphasis is beneficial since both biomedical and acoustic signals are highly dependent on information present in the low frequency range~\cite{Mel-low-frequency}. 

Like speech, EHG signals are non stationary (their frequency content changes over time)\cite{2012Dataset} when viewed over long durations. MFCCs are designed to capture such time varying spectral features effectively. The emphasis on lower frequencies aligns with physiological observations of uterine activity, where low frequency components are often more informative, especially in late pregnancy or during labor.

MFCCs can be thought of as the construction of a compact signature of the EHG signal in the form of a coefficient vector. This provides a low dimensional representation that emphasizes the physiologically relevant low frequency components associated with PTB while suppressing irrelevant details. Since MFCCs operate in the cepstral domain (log-power spectrum), they are relatively robust to such noise. They can help separate meaningful signal components from interference, improving the reliability of feature extraction.

Equation \eqref{eq:F_mel} shows the relationship between Mel and Hertz frequencies and how the Mel scale shrinks the frequency axis in a perceptually valuable way, where \textit{f} is sampling frequency in Hz.

\begin{equation}
    \label{eq:F_mel} 
F_{\text{mel}} = 2595 \cdot \log_{10}\left(1 + \frac{f}{700}\right)
\end{equation}

\subsubsection{Wavelet coefficients extraction}

Wavelet coefficients are numerical representations derived from the wavelet transform, which decomposes a signal into components at multiple scales and positions. Unlike the Fourier transform, which provides only frequency information, the wavelet transform captures both time and frequency localization~\cite{wavelet-theory}. This is particularly useful for analyzing non stationary biomedical signals, such as EHG, where the signal characteristics vary over time.

The Discrete Wavelet Transform (DWT) applies a series of high pass and low pass filters to the signal, separating it into approximation (low frequency) and detail (high frequency) components at multiple levels. The resulting wavelet coefficients summarize how the signal behaves at different time–frequency resolutions.

EHG signals exhibit transient events, such as uterine contractions, which are better captured in the time-frequency domain. Wavelet coefficients can isolate these transient patterns and highlight changes in energy over time. Features such as the energy or entropy of wavelet sub bands are often used for classification tasks, as they effectively capture signal characteristics relevant to distinguishing between term and preterm activities. Wavelets are particularly effective at capturing the non white, structured variabilities of these transients, and effects that can be further enhanced when KLT suppresses the underlying white noise floor.

Choosing the right mother wavelet $\psi(t)$ is crucial to the effectiveness of the Wavelet Transform's (WT)  since it affects the precision of the time-frequency representation (TFR) of the original signal that is produced \cite{mother-wavelet}. The wavelet is typically expressed in its scaled and translated form:   
\begin{equation}
\psi_{s,u}(t) = \frac{1}{\sqrt{s}} \, \psi\left( \frac{t - u}{s} \right)
\end{equation}
where $s$ is the scale parameter that controls frequency resolution, and $u$ is the translation parameter that determines time localization.

In this work, the Daubechies 8 (db8) wavelet was selected because it effectively captures both the smooth low frequency uterine activity and the sharp transient changes, making it more suitable than shorter Daubechies wavelets. This property makes it particularly appropriate for analyzing the nonstationary nature of EHG signals \cite{BATISTA2016178}. Accordingly, the DWT was implemented using the db8 wavelet with a five level decomposition to analyze the EHG signals.

For each channel of the EHG signals, the five level wavelet decomposition produces five sets of detail coefficients (D1-D5) and one set of approximation coefficients (A5), yielding six sub bands in total. From each sub band, six statistical features were extracted from the wavelet coefficients to characterize the signal's time-frequency properties. This results in 6 features $\times$ 6 sub bands, which is 36 features per channel. These features are computed from the coefficients within each sub band, where $c[i]$ represents the $i$-th wavelet coefficient and $N$ is the total number of coefficients in that sub band.

The extracted features include:
\begin{enumerate}
\item Mean
\begin{equation}
\mu = \frac{1}{N} \sum_{i=1}^{N} c[i]
\end{equation}

\item Variance
\begin{equation}
\sigma^2 = \frac{1}{N} \sum_{i=1}^{N} (c[i] - \mu)^2
\end{equation}
\item Energy
 \begin{equation}
 E = \sum_{i=1}^{N} c[i]^2
\end{equation}
\item Absolute Sum
\begin{equation}
S_{\text{abs}} = \sum_{i=1}^{N} |c[i]|
\end{equation}
\item Skewness
 \begin{equation}
 \text{Skew} = \frac{\frac{1}{N} \sum_{i=1}^{N} (c[i] - \mu)^3}{\left( \sqrt{\frac{1}{N} \sum_{i=1}^{N} (c[i] - \mu)^2} \right)^3}
\end{equation}
\item Kurtosis
\begin{equation}
\text{Kurt} = \frac{\frac{1}{N} \sum_{i=1}^{N} (c[i] - \mu)^4}{\left( \frac{1}{N} \sum_{i=1}^{N} (c[i] - \mu)^2 \right)^2} - 3
\end{equation}
\end{enumerate}

These statistical measures are chosen because they summarize key characteristics of the signal in each sub band, such as average activity, variability, energy distribution, and the shape of the coefficient distribution, helping to capture patterns associated with uterine behavior. By examining a range of statistics, including higher order moments, wavelet features can capture not only energy and variability but also transient dynamics and subtle changes that are crucial for distinguishing term from preterm signals. This approach enables the effective representation of essential patterns in both the time and frequency domains, providing discriminative information for classification.

\subsection{Model Selection}

Given the characteristics of the dataset and the relatively limited size of available labeled EHG recordings, classical ML techniques were employed for classifying signal segments. These standard ML methods were chosen because, with limited data, more complex approaches such as DL can easily overfit. Coupled with the efficient denoising and feature selection strategies described in the previous sections, these computationally simpler methods are sufficient to capture relevant patterns. A set of supervised binary classification models was selected based on their proven performance in prior studies focused on PTB prediction and similar biomedical signal classification tasks \cite{GarciaCasado2018, XU2022103231,IEEE_Access_2024}. The selected models include:

\begin{itemize}
    \item QDA - Quadratic Discriminant Analysis
    \item LR - Logistic Regression with L2 regularization
    \item SVM - Support Vector Machine with a linear kernel and regularization parameter \( C = 1.0 \)
    \item DT - Decision Tree Classifier with a maximum depth of 100
    \item RF - Random Forest Classifier with 100 estimators and maximum depth of 10
    \item GB - Gradient Boosting Classifier with 100 estimators and a learning rate of 0.1
    \item MLP - Multi-Layer Perceptron Classifier with one hidden layer of 100 units and ReLU activation
    \item CB - CatBoost Classifier with 100 iterations and a learning rate of 0.1
\end{itemize}

All models were implemented using the Scikit learn (sklearn) library. These models were chosen to reflect a diverse range of learning paradigms and to enable comparative analysis of their predictive capabilities on the EHG dataset.

\subsection{Evaluation Protocol and Metrics}
\label{sec:evaluation}

Segmentation defines the sample unit as described in Section~\ref{sec:segmentation}. All reported results are computed at the segment level (contraction/dummy intervals or fixed 3-minute windows), where each segment inherits the corresponding record label (term vs.\ preterm).

To evaluate each binary classifier, we report Classification Accuracy, Precision, Recall, F1-score, and the Area Under the Curve (AUC). Let $TP$, $TN$, $FP$, and $FN$ denote the counts of correctly predicted preterm segments, correctly predicted term segments, term segments misclassified as preterm, and preterm segments misclassified as term, respectively. The metrics are computed as:
\begin{equation}
\text{Accuracy} = \frac{TP + TN}{TP + TN + FP + FN}
\end{equation}
\begin{equation}
\text{Precision} = \frac{TP}{TP + FP}
\end{equation}
\begin{equation}
\text{Recall} = \frac{TP}{TP + FN}
\end{equation}
\begin{equation}
\text{F1-score} = 2 \times \frac{\text{Precision} \times \text{Recall}}{\text{Precision} + \text{Recall}}
\end{equation}
\begin{equation}
\text{AUC} = \int_{0}^{1} \text{TPR}(\text{FPR})\, d(\text{FPR})
\end{equation}

AUC summarizes the discriminative ability of the classifier across all decision thresholds and equals the area under the Receiver Operating Characteristic (ROC) curve.

\textit{Iterative balanced sampling:} To address class imbalance for the TPEHGT contraction--dummy interval setting (Section~\ref{subsec:seg_contraction_dummy}), we employ an iterative balanced sampling strategy:
\begin{itemize}
    \item \textbf{Minority class:} all available preterm segments are used (94 total: 47 contraction + 47 dummy).
    \item \textbf{Majority class:} an equal number of term segments (94) are randomly selected from the available pool (106 total: 53 contraction + 53 dummy).
\end{itemize}
This balancing is repeated for twenty independent iterations using different random seeds for the term undersampling. Reported results are averaged across the twenty iterations.

\textit{Cross validation:} Within each balanced subset, we perform five fold stratified cross validation at the segment level to preserve the class distribution across folds. The cross-validation split is generated using a fixed random seed for reproducibility, while the majority class subsampling varies across iterations.

\section{Results and Discussion}
\label{sec:results}

This section presents the classification results and key findings from our experiments. First, Section~\ref{sec:baseline} reports baseline performance on the TPEHGT dataset, including results with and without KLT based denoising. Section~\ref{sec:ablation} provides ablation studies to analyze the contribution of different factors. In particular, Section~\ref{subsec:impact_klt} quantifies the performance gain from applying KLT, after which all subsequent ablation results are reported using KLT transformed signals only. Section~\ref{subsec:with_toco} and Section~\ref{subsec:without_contraction} then evaluate the influence of fixed interval segmentation with and without TOCO signal, respectively. Next, Section~\ref{sec:generalization} assesses the generalization of the proposed approach on the TPEHG database. Finally, Section~\ref{sec:comparison} compares our best performing model against existing methods in the literature.

\subsection{Performance Evaluation on TPEHGT dataset}
\label{sec:baseline}

The classification methods discussed earlier were implemented using the TPEHGT dataset~\cite{2018Dataset}. We report results using the contraction-dummy interval samples defined in Sec.~\ref{subsec:seg_contraction_dummy}, yielding 188 segment samples in total. Feature extraction was carried out using 20 MFCCs, along with six statistics of db8 wavelet coefficients extracted from five detail sub bands (D1–D5) and one approximation sub band (A5) per channel, resulting in 36 wavelet based features (6 statistics $\times$ 6 sub bands), plus the PA of the normalized power spectrum. Combined with 20 MFCCs, this yields a total of 57 features per channel. Considering three EHG channels, the total feature dimension is $57 \times 3 = 171$.

To evaluate classification performance, we tested eight classical ML models. Performance is reported using the evaluation protocol in Section~\ref{sec:evaluation}, including iterative balanced sampling and five fold stratified cross validation. Model performance was compared on both raw filtered signals and their KLT versions, which denoise and decorrelate the signals to improve feature quality. The results are summarized in Table~\ref{tab:datasets}.

\begin{table}[tb]
\centering
\caption{Classification performance of eight models on the 2018 TPEHGT dataset using contraction-dummy interval sampling. Results are reported for both the original signals and those processed with the KLT.}
\resizebox{\columnwidth}{!}{%
\begin{tabular}{llccccc}
\toprule
\textbf{Signal} & \textbf{Model} & \textbf{Accuracy} & \textbf{Precision} & \textbf{Recall} & \textbf{F1 Score} & \textbf{AUC} \\
\midrule
\multirow{8}{*}{\shortstack{Without\\KLT}}
  & QDA & 89.05 & 93.88 & 83.82 & 88.37 & 0.9281 \\
  & LR  & 89.69 & 91.30 & 88.36 & 89.46 & 0.9427 \\
  & SVM & 88.04 & 90.09 & 86.24 & 87.75 & 0.9306 \\
  & DT  & 87.78 & 87.59 & 88.64 & 87.90 & 0.8779 \\
  & RF  & \textbf{92.14} & \textbf{94.55} & 89.73 & 91.97 & 0.9662 \\
  & GB  & 91.41 & 93.71 & 89.10 & 91.18 & 0.9659 \\
  & MLP & 88.91 & 90.58 & 87.42 & 88.49 & 0.9256 \\
  & CB  & 91.97 & 92.87 & \textbf{91.42} & \textbf{91.99} & \textbf{0.9717} \\
\midrule
\multirow{8}{*}{\shortstack{With\\KLT}}
  & QDA & \textbf{94.00} & \textbf{97.69} & 90.25 & \textbf{93.75} & 0.9514 \\
  & LR  & 90.92 & 90.41 & \textbf{92.05} & 91.03 & 0.9483 \\
  & SVM & 90.28 & 90.10 & 91.10 & 90.34 & 0.9411 \\
  & DT  & 86.11 & 87.11 & 85.63 & 86.08 & 0.8609 \\
  & RF  & 93.10 & 96.04 & 90.15 & 92.90 & \textbf{0.9746} \\
  & GB  & 91.20 & 95.13 & 87.06 & 90.76 & 0.9716 \\
  & MLP & 90.29 & 91.31 & 89.52 & 90.19 & 0.9359 \\
  & CB  & 92.99 & 95.56 & 90.47 & 92.81 & \textbf{0.9746} \\
\bottomrule
\end{tabular}
}
\label{tab:datasets}
\end{table}

The results shown in Table \ref{tab:datasets} reflect model performance when EHG signals were segmented based on contraction and non contraction (dummy) intervals, which simulate annotated contraction periods and the intervening rest periods. This segmentation approach inherently prioritizes physiologically relevant signal segments, which likely explains the overall strong performance across models.  

The models achieved good classification accuracy, with most scoring between 86–94\%. Ensemble methods, particularly RF and GB variants like CB, consistently delivered the strongest performance across all evaluation metrics, reaching the highest accuracy levels (92–93\%) and AUC scores (0.96–0.97) by effectively combining multiple weak learners and capturing complex, nonlinear relationships in the data. Single model approaches such as QDA, LR, SVM, DT, and MLP performed slightly lower but still produced robust results. The relatively small performance gains of ensemble methods after denoising highlight their inherent robustness to noise, making them well suited for classifying subtle physiological signals such as EHG.

QDA, on the other hand, exhibited the most substantial improvement from denoising, ultimately becoming the highest performing model post KLT. After denoising, it achieved an accuracy of 94.00\% and a precision of 97.69\%, significantly surpassing its original performance. This aligns with the expectation that models more sensitive to noise benefit most from KLT enhancement, whereas noise robust algorithms show smaller gains, further validating the effectiveness of our denoising approach.

\subsection{Ablation Studies}
\label{sec:ablation}

\subsubsection{The impact of KLT}
\label{subsec:impact_klt}

In the proposed pipeline, denoising was performed using the KLT to suppress residual noise and overlapping signal components that persist after band pass filtering. To assess its effectiveness, we evaluated classification performance both with and without KLT, as shown in Table \ref{tab:datasets}.

Across most classifiers, KLT improved performance, particularly in terms of accuracy, F1-score, and AUC. For example, the QDA accuracy increased from 89.05\% to 94\%, with a notable F1-score gain from 88.37 to 93.75. Similarly, the RF and CB classifiers achieved AUC values of 0.9746, outperforming their non KLT counterparts.

These improvements are consistent with the theoretical properties of KLT, which preserve signal energy while discarding noise dominated components. This effect is particularly beneficial for noise sensitive models like QDA and margin based classifiers like SVM, which rely on clearer class boundaries for accurate separation. In contrast, ensemble methods such as RF, GB, and CB already robust to noisy inputs due to bagging or boosting, showed relatively modest gains, though they still delivered the strongest overall performance.

Overall, these findings confirm that KLT enhances the discriminative features of EHG signals and boosts classification reliability, especially for models sensitive to distributional assumptions. Therefore, KLT was retained in all subsequent ablation studies and validation experiments.

\subsubsection{Performance with TOCO signals}
\label{subsec:with_toco}
In real time PTB detection, identifying contraction intervals from EHG signals using TOCO is not always practical. While some studies have attempted to detect contractions directly from EHG signals~\cite{Chen2024}, this remains a challenging task. To evaluate the potential benefits of including TOCO, we adopt the fixed 3-minute non overlapping segmentation described in Section~\ref{subsec:seg_fixed3min}, which simulates a blind real time monitoring setup. For this setting, both EHG channels and the TOCO signal were used during feature extraction. For TPEHGT, this produces 260 segment samples (130 term + 130 preterm) under the fixed window setup. Feature extraction followed the same pipeline as in the contraction based setup, using twenty MFCCs, six statistical features from db8 wavelet coefficients across five detail sub bands (D1–D5) and one approximation sub-band (A5), and the PA of the normalized power spectrum, resulting in a total of 57 features per channel.

We selected 3-minute non overlapping windows as a balanced choice between physiology and signal analysis. Since uterine contractions typically last 1–2 minutes, a 3-minute interval provides sufficient coverage of complete contraction events while also capturing the surrounding baseline activity \cite{ncc_uterine_activity, McEvoy2022}. This window length balances contextual coverage, avoids excessive signal fragmentation, and supports computational efficiency for real time applications.

Table~\ref{tab:datasets_3min_2018_with_TOCO} presents the classification results for the TPEHGT dataset using fixed 3-minute intervals, where both EHG and TOCO signals were included during feature extraction. 

\begin{table}[tb]
\centering
\caption{Classification performance of eight models on the 2018 TPEHGT dataset (with TOCO signal) using fixed 3-minute interval sampling. Results are shown for those transformed with the KLT. Evaluation metrics include Accuracy, Precision, Recall, F1 Score, and AUC.}
\resizebox{\columnwidth}{!}{%
\begin{tabular}{lccccc}
\toprule
\textbf{Model} & \textbf{Accuracy} & \textbf{Precision} & \textbf{Recall} & \textbf{F1 Score} & \textbf{AUC} \\
\midrule
  QDA & 95.81 & 98.12 & 93.50 & 95.66 & 0.9407 \\
  LR  & 92.86 & 92.37 & 93.75 & 92.96 & 0.9524 \\
  SVM & 91.35 & 91.03 & 92.12 & 91.45 & 0.9447 \\
  DT  & 90.42 & 90.39 & 90.77 & 90.32 & 0.9041 \\
  RF  & 96.35 & \textbf{98.86} & 93.84 & 96.20 & 0.9981 \\
  GB  & 96.70 & 98.52 & 94.85 & 96.55 & 0.9969 \\
  MLP & 87.50 & 89.32 & 85.79 & 87.03 & 0.9188 \\
  CB  & \textbf{97.28} & 98.78 & \textbf{95.78} & \textbf{97.21} & \textbf{0.9988} \\
\bottomrule
\end{tabular}
}
\label{tab:datasets_3min_2018_with_TOCO}
\end{table}

In the fixed interval setup, all models demonstrated strong performance, with classification accuracies ranging from 87\% to 97\%. Ensemble methods, particularly RF, GB and CB, achieved the highest performance across all metrics. CB reached the highest accuracy of 97.28\% and an AUC of 0.9988 on the KLT transformed signals, indicating that fixed 3-minute windows can capture enough discriminative signal characteristics for effective PTB prediction, even in the absence of explicit contraction boundaries.

Comparing Tables~\ref{tab:datasets} and \ref{tab:datasets_3min_2018_with_TOCO}, we observe a trade off between contraction based and fixed interval segmentation: some models excel in one setup but not the other. It is important to note that direct comparison is not meaningful because the datasets differ in both sample size and segmentation strategy. The contraction based dataset contains 188 samples, while the fixed size window dataset includes 260 samples. Despite these differences, the aim is not to rank the segmentation strategies but to highlight the overall versatility of the proposed approach. The combination of KLT based denoising, which compensates for noise vulnerability in specific models, and contextual feature selection, which emphasizes physiologically relevant patterns, ensures that model performance remains consistently high across different scenarios. This demonstrates the method's adaptability to diverse real world conditions, ranging from fully monitored clinical settings to real time or low resource applications.

\subsubsection{Performance without TOCO signals}
\label{subsec:without_contraction}
In real time PTB detection, access to TOCO signals may not always be feasible, particularly in low resource or wearable monitoring setups. To simulate such scenarios, we use the same fixed 3-minute segmentation described in Section~\ref{subsec:seg_fixed3min} while excluding TOCO features and using only the EHG channels. This approach avoids reliance on external contraction measurements and tests the model's robustness in practical, TOCO independent conditions.

Table~\ref{tab:datasets_3min_2018} summarizes the classification results for the TPEHGT dataset without TOCO signals. Overall, the models achieved accuracies ranging from 88\% to 96\%, which is only slightly lower than the with TOCO setup (Table~\ref{tab:datasets_3min_2018_with_TOCO}, 87--97\%). Similar to the with TOCO case, ensemble methods (RF, GB, CB) consistently outperformed classical models, with GB achieving the highest accuracy of 96.82\% and  a strong AUC of 0.9953 on the KLT denoised signals.

\begin{table}[tb]
\centering
\caption{Classification performance of eight models on the 2018 TPEHGT dataset (without TOCO signal) using fixed 3-minute interval sampling. Results are shown for those transformed with the KLT. Evaluation metrics include Accuracy, Precision, Recall, F1 Score, and AUC.}
\resizebox{\columnwidth}{!}{%
\begin{tabular}{llccccc}
\toprule
\textbf{Model} & \textbf{Accuracy} & \textbf{Precision} & \textbf{Recall} & \textbf{F1 Score} & \textbf{AUC} \\
\midrule
  QDA & 94.43 & 97.14 & 91.71 & 94.23 & 0.9270 \\
  LR  & 88.90 & 90.68 & 87.42 & 88.58 & 0.9422 \\
  SVM & 88.50 & 89.84 & 87.52 & 88.33 & 0.9257 \\
  DT  & 90.57 & 91.09 & 90.11 & 90.38 & 0.9056 \\
  RF  & 96.06 & 98.45 & 93.67 & 95.92 & 0.9971 \\
  GB  & \textbf{96.82} & 98.04 & \textbf{95.58} & 96.72 & 0.9953 \\
  MLP & 92.03 & 93.27 & 91.03 & 91.96 & 0.9448 \\
  CB  & 96.79 & \textbf{98.47} & 95.13 & \textbf{96.70} & \textbf{0.9984} \\
\bottomrule
\end{tabular}
}
\label{tab:datasets_3min_2018}
\end{table}

Compared to the with TOCO setup, the performance differences were marginal. For example, CB achieved 97.28\% accuracy with TOCO versus 96.79\% without, indicating that TOCO contributes only modest improvements when robust EHG based features are already utilized. Importantly, these findings demonstrate that fixed 3-minute EHG intervals capture sufficient discriminative patterns for reliable classification, even without auxiliary contraction measurements. This highlights the practicality of the proposed approach in real world scenarios, where TOCO signals may be unavailable or unreliable, such as in ambulatory or wearable monitoring applications.

\subsection{EVALUATION ON THE TPEHG DATASET}
\label{sec:generalization}

To further evaluate the proposed pipeline on an additional dataset, we used the 2012 TPEHG dataset. Since this dataset lacks TOCO signals, we apply the fixed 3-minute non overlapping segmentation described in Section~\ref{subsec:seg_fixed3min} and extract features from EHG channels only. The original dataset contained 300, thirty minute recordings (38 preterm and 262 term). Through our balanced sampling strategy, we generated 760 segments (380 preterm and 380 term) per training iteration. From each segmented signal, we extracted 57 features per channel as previously described. The corresponding performance metrics are presented in Table~\ref{tab:datasets_3min_2012}.

\begin{table}[tb]
\centering
\caption{Classification results for eight models applied to the 2012 TPEHG dataset using fixed 3-minute interval sampling. KLT processed signal variants are evaluated across Accuracy, Precision, Recall, F1 Score, and AUC metrics.}
\resizebox{\columnwidth}{!}{%
\begin{tabular}{llccccc}
\toprule
\textbf{Model} & \textbf{Accuracy} & \textbf{Precision} & \textbf{Recall} & \textbf{F1 Score} & \textbf{AUC} \\
\midrule
  QDA & 86.97 & \textbf{92.58} & 80.56 & 85.96 & 0.9196 \\
  LR  & 70.85 & 70.22 & 72.58 & 71.25 & 0.7626 \\
  SVM & 70.56 & 69.52 & 73.39 & 71.30 & 0.7498 \\
  DT  & 70.91 & 70.79 & 71.57 & 71.09 & 0.7091 \\
  RF  & 88.07 & 86.57 & 90.41 & 88.33 & 0.9550 \\
  GB  & 85.42 & 82.96 & 89.38 & 85.96 & 0.9292 \\
  MLP & 79.49 & 78.50 & 82.22 & 79.85 & 0.8568 \\
  CB  & \textbf{90.31} & 87.79 & \textbf{93.84} & \textbf{90.63} & \textbf{0.9705} \\
\bottomrule
\end{tabular}
}
\label{tab:datasets_3min_2012}
\end{table}

When evaluated on the larger 2012 TPEHG dataset, the classification results were slightly lower compared to those obtained on the 2018 dataset, despite using the same feature extraction pipeline and segmentation strategy. This decline is expected given the increased variability and the strong class imbalance in the original dataset (38 preterm vs. 262 term). Although balanced sampling was applied to generate 720 equal segments per iteration, the limited preterm pool still means that variability in the selected subsets can impact the generalizability of learned patterns. Nevertheless, ensemble models like CB and RF continued to deliver good performance, achieving accuracies of 90.31\% and 88.07\%, respectively, with AUCs above 0.95 after KLT denoising.

These results also highlight the limitations of the available datasets and the challenges posed by sampling strategies. Despite balancing the classes during training, the original imbalance in subject distribution, particularly the low number of preterm cases, can still influence model generalization. This limitation is further emphasized in Section~\ref{sec:comparison}, where we compare our approach to existing models. The analysis reveals that improper oversampling strategies employed in prior works have significantly inflated reported performance, often resulting in overly inflated accuracy metrics that do not accurately reflect real world deployment conditions.

\begin{figure}[tb]
    \centering
    \includegraphics[width=\linewidth]{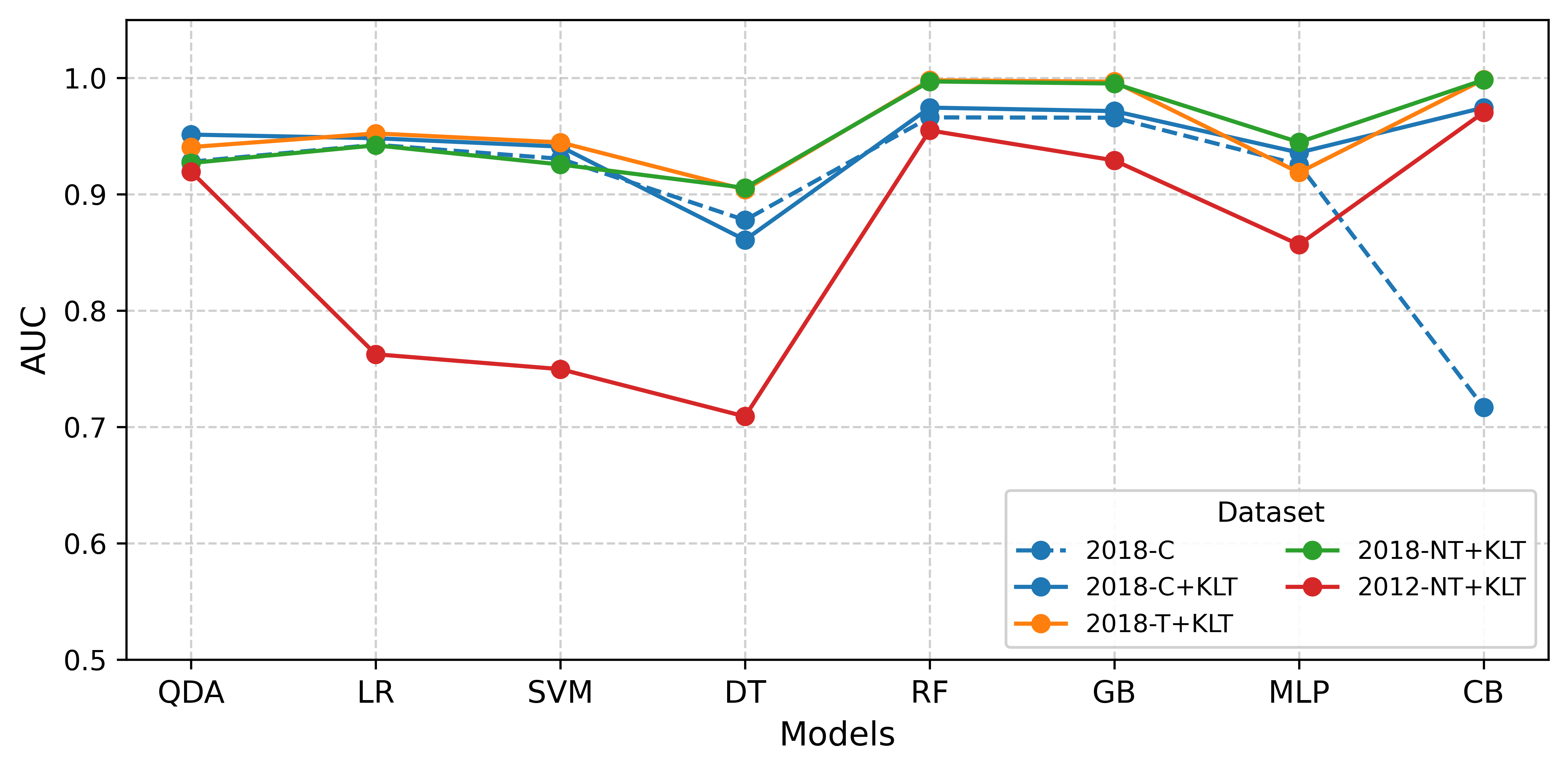}
    \caption{Model-wise classification AUCs on the 2012 TPEHG and 2018 TPEHGT datasets. 
    Legend entries indicate segmentation type and KLT usage: 
    C = contraction-based (2018), T = fixed 3-min windows with TOCO (2018), NT = fixed 3-min windows without TOCO (2012, 2018). 
    Dashed lines = original signals, solid lines = KLT-denoised signals.}
    \label{fig:comparison}
\end{figure}

Figure~\ref{fig:comparison} summarizes the classification performance across different experimental setups. For the 2018 contraction interval dataset, results are shown for both original signals and KLT denoised signals, clearly highlighting the performance gains achieved through KLT. In contrast, for the 2018 fixed interval experiments (with and without TOCO) and the 2012 fixed interval dataset (without TOCO), only KLT enhanced results are reported, since KLT was consistently found to improve accuracy. Overall, the figure demonstrates that KLT denoising strengthens model performance, while the evaluation across multiple segmentation strategies and datasets reinforces the robustness and adaptability of the proposed approach in diverse clinical conditions.

\subsection{Comparison with existing models}
\label{sec:comparison}
Building upon the methodological concerns raised earlier regarding inflated performance metrics from improper oversampling, our comparative analysis adopts a rigorous approach to benchmark evaluation. We restrict our comparison to studies that have either been methodologically validated by Vandewiele et al. \cite{VANDEWIELE2021101987} or that explicitly document appropriate sampling strategies in their original publications. Table \ref{tab:comparison_reported_tuned_metric} presents this curated comparison, including both originally reported results and, where available, the methodologically corrected results from Vandewiele et al. In addition, we incorporate recent classical ML studies that report reproducible methodologies and clinically interpretable outcomes. In addition to the classifiers explored in this study, results from prior works using AdaBoost (AB) are also included for completeness.

Notably, we exclude DL based approaches from this comparison due to fundamental limitations of their substantial data requirements, which cannot be met by currently available EHG datasets. Additionally, there is an inherent lack of interpretability in automated feature extraction, which prevents the provision of meaningful physiological insights into the mechanisms of preterm birth \cite{Jager2024}.



\begin{table}[tb]
\centering
\caption{Comparison of reported and corrected AUC values from previous studies.}
\label{tab:comparison_reported_tuned_metric}
\resizebox{\columnwidth}{!}{%
\begin{tabular}{c c c c c}
\toprule
Study & Classifier & Metric & {\shortstack{Uncorrected\\Results}} & {\shortstack{Corrected\\Results}} \\
\midrule
\cite{Fergus2013, VANDEWIELE2021101987} & SVM & AUC & 0.9200 & 0.6075 \\
\cite{Ren2015, VANDEWIELE2021101987} & AB & AUC & 0.9860 & 0.5754 \\
\cite{ref30, VANDEWIELE2021101987} & RF & AUC & 0.9400 & 0.5231 \\
\cite{Ahmed2016_fuzzyEHG, VANDEWIELE2021101987} & SVM & AUC & 0.9900 & 0.5604 \\
\cite{2018Dataset, VANDEWIELE2021101987} & QDA & AUC & 0.9944 & 0.6386 \\
\cite{Peng2020, VANDEWIELE2021101987} & RF & AUC & 0.8880 & 0.5185 \\

\midrule
\cite{Frontier_2024} & CB & AUC & NA & 0.7000 \\
\cite{LOU2022103587}& Bayes & AUC & NA & 0.8400 \\
\cite{Jager2024} & SVM & AUC & NA & 0.8600 \\
\cite{10.3389/fendo.2022.1035615} & SVM & AUC & NA & 0.8900 \\
\cite{73_Jager2020-in}& QDA& AUC& NA&0.9030\\
\cite{78_Xu2022-ge}& HVG + SVM& AUC& NA & 0.9700\\

\midrule
Proposed & CB & AUC & NA & 0.9988 \\
\bottomrule
\end{tabular}%
}
\end{table}

PTB detection is a critical and sensitive clinical task; yet, much of the prior literature has reported overly optimistic results, portraying PTB prediction as a solved problem. In reality, this is far from the truth. As shown in Table~\ref{tab:comparison_reported_tuned_metric}, many earlier studies relied on flawed sampling strategies, particularly improper oversampling and data leakage, leading to dramatically inflated performance metrics. For instance, several reported AUCs exceeding 90\% dropped on average by more than 25 percentage points, in some cases falling below 65\% when evaluated using proper methodology. This striking discrepancy highlights the extent of overestimation in prior work and underscores the urgent need to reopen and rigorously reevaluate the PTB prediction problem. Notably, the limitations of current datasets continue to be a significant barrier to the development of generalizable and reliable models. This study represents an essential first step in that re-evaluation, while also calling for more extensive datasets and methodological rigor to add credibility to AI based PTB research. In doing so, it aims to build clinical trust by prioritizing explainability, transparency, and interoperability in model design.

Our experiments using the comparatively larger and more variable 2012 TPEHG dataset clearly demonstrate how classification performance shifts when dataset variability increases. Models that performed well on smaller or more uniform datasets often struggle when exposed to real world heterogeneity. This underlines the importance of features that go beyond surface level patterns and instead capture the underlying physiological structures of EHG signals. 

Our results show that frequency domain features, especially the combination of MFCCs, wavelet based statistical features, and the PA of the normalized power spectrum, provide a more reliable representation of uterine activity. Each of these features contributes uniquely: MFCCs effectively capture time varying spectral content with an emphasis on lower frequencies, aligning with known uterine activity patterns. Wavelet coefficients provide multi resolution time–frequency insights, useful for identifying transient contraction events. Finally, the PA captures dominant frequency behavior linked to gestational progression. Together, these features form a robust representation of EHG signals, helping bridge the gap between physiological relevance and model performance. As such, future efforts must prioritize physiologically grounded features and rigorous validation to ensure true progress in the field.

\section{Conclusion}
\label{conclusion}

Preterm birth (PTB) affects approximately 10\% of live births and remains the leading cause of death among children under five. Survivors often face lifelong challenges, including developmental delays, chronic diseases, and sensory impairments. Early and accurate prediction of PTB is therefore essential for improving neonatal outcomes and guiding timely interventions. This study proposed a robust and reproducible machine learning (ML) pipeline for PTB prediction using EHG signals, addressing key limitations in prior research, particularly issues with improper oversampling strategies and limited focus on frequency domain features.

A central contribution of this work was the integration of the Karhunen Loève Transform (KLT), which effectively suppressed residual and overlapping noise components that persist after band pass filtering. By improving the signal to noise ratio, KLT enhanced the quality of the extracted features and boosted model performance. For example, in the contraction based segmentation of the 2018 TPEHGT dataset, Quadratic Discriminant Analysis (QDA) accuracy improved from 89.05\% to 94.00\%, with a notable F1-score gain from 88.37 to 93.75. In the fixed 3-minute setup with TOCO signals, the CatBoost (CB) classifier achieved the highest performance, reaching 97.28\% accuracy and an AUC of 0.9988. These results confirm that denoising substantially benefits noise sensitive models, while ensemble classifiers such as RF, GB, and CB maintain strong robustness across all setups.

Feature extraction in the proposed pipeline was designed to capture the physiological characteristics of uterine activity, and was applied consistently in both the original and KLT denoised cases. We employed Mel frequency cepstral coefficients (MFCCs) to compress the signal’s spectral envelope using a perceptually motivated filter bank with emphasis on low frequency components; statistical descriptors of wavelet coefficients across multiple decomposition levels to characterize localized time–frequency behavior; and the peak amplitude (PA) of the normalized power spectrum to capture dominant frequency content. This combination of cepstral, time–frequency, and spectral peak features provided a compact yet expressive representation of uterine dynamics, enabling robust discrimination between term and preterm cases.

An additional contribution of this work is the direct comparison of prediction performance with and without the inclusion of TOCO signals. Our findings indicate that TOCO does not provide a significant improvement over EHG alone, underscoring the latter’s sufficiency as a standalone modality for PTB prediction. This result is clinically important, as it supports the use of EHG based monitoring in low resource settings where TOCO devices may not be available or feasible.

Despite these advancements, essential limitations remain, most notably the reliance on publicly available datasets with limited samples of PTB and constrained variability. These factors limit the generalizability of the results to broader populations. This study also highlights a critical issue in existing literature: PTB detection has often been treated as a closed problem, with artificially inflated results produced by flawed sampling strategies and unrealistic evaluation setups. However, as demonstrated in our experiments with a more variable dataset, predictive performance can shift substantially under more realistic conditions. This underscores that PTB prediction is a very open and unresolved problem. Future work should prioritize expanding dataset diversity and rigor, while advancing explainable and physiologically grounded features to support clinical trust and adoption. By reframing PTB detection as an ongoing challenge that demands broader and more representative analysis, this study provides an essential step toward developing scalable, reliable, and clinically relevant tools, particularly in low resource settings where such interventions are most needed.

\bibliographystyle{ieeetr}
\bibliography{paper-refs}

\begin{IEEEbiography}[{\includegraphics[width=1in,height=1.25in,clip,keepaspectratio]{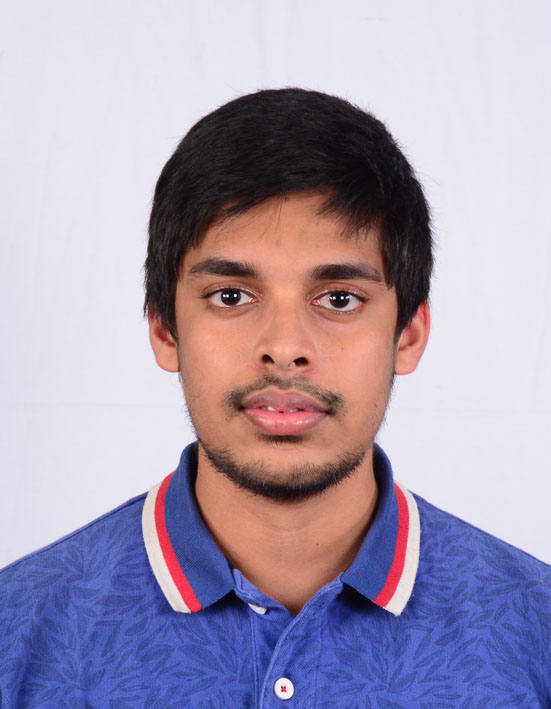}}]{Senith Jayakody}
received the B.Sc. (Eng.) degree in electrical and electronic engineering from the University of Peradeniya, Sri Lanka. He is currently a Research Assistant with the Multidisciplinary AI Research Centre (MARC), University of Peradeniya. He has published papers in multiple IEEE conferences. His research interests include machine learning, signal processing, and computer vision for biomedical and real-world sensing.
\end{IEEEbiography}

\begin{IEEEbiography}[{\includegraphics[width=1in,height=1.25in,clip,keepaspectratio]{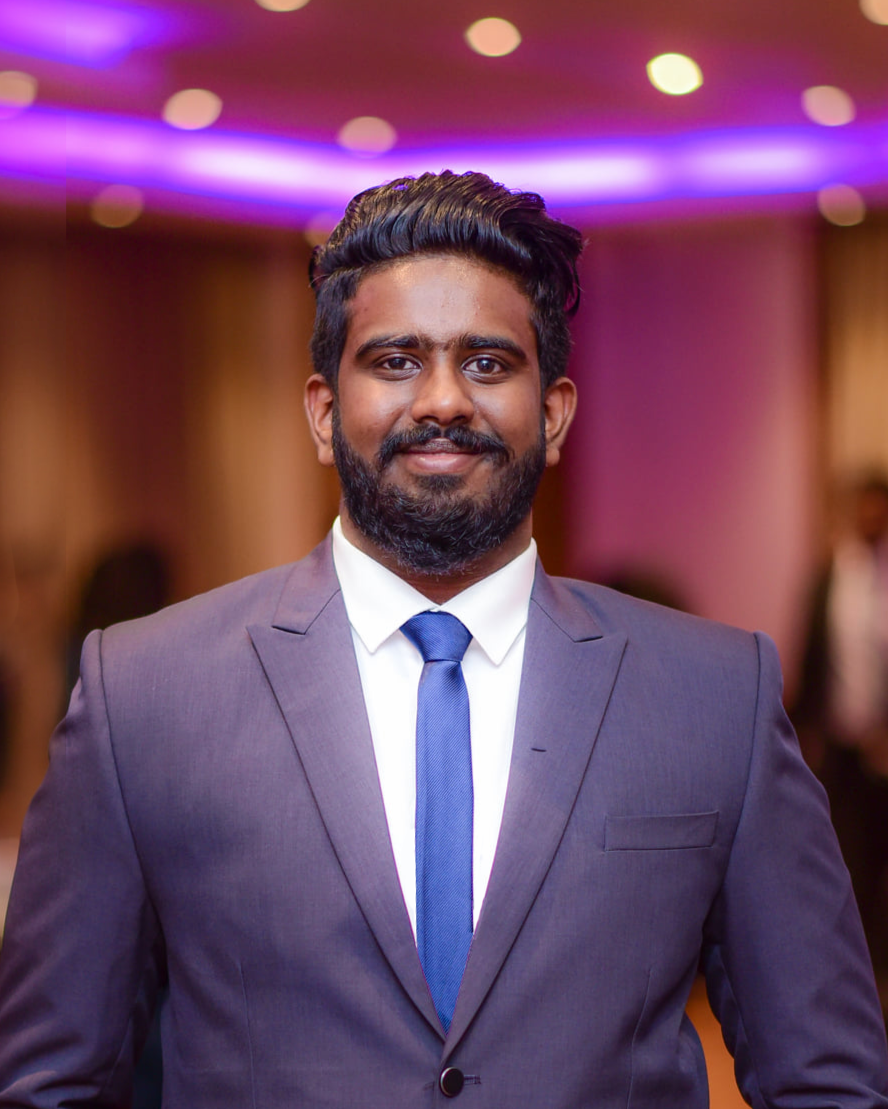}}]{Kalana Jayasooriya} received the B.Sc. degree in engineering from the University of Peradeniya, Sri Lanka, in 2024. He is currently a Research Assistant with the Multidisciplinary AI Research Center, University of Peradeniya, where he works on signal processing, machine learning, and AI-driven modeling for biomedical and interdisciplinary applications.

\end{IEEEbiography}



\begin{IEEEbiography}[{\includegraphics[width=1in,height=1.25in,clip,keepaspectratio]{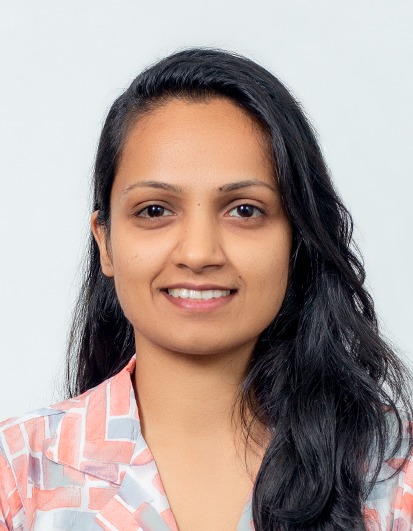}}]{SASHINI LIYANAGE} received the B.Sc. degree in engineering from the University of Peradeniya, Sri Lanka, in 2023. At the time of this work, she was a Research Assistant at the Multidisciplinary AI Research Center, University of Peradeniya. She is currently pursuing the Ph.D. degree at Princeton University, USA, with a focus on natural language processing for medical reasoning. Her research interests include machine learning, signal processing, and natural language processing for biomedical AI applications.
\end{IEEEbiography}

\begin{IEEEbiography}[{\includegraphics[width=1in,height=1.25in,clip,keepaspectratio]{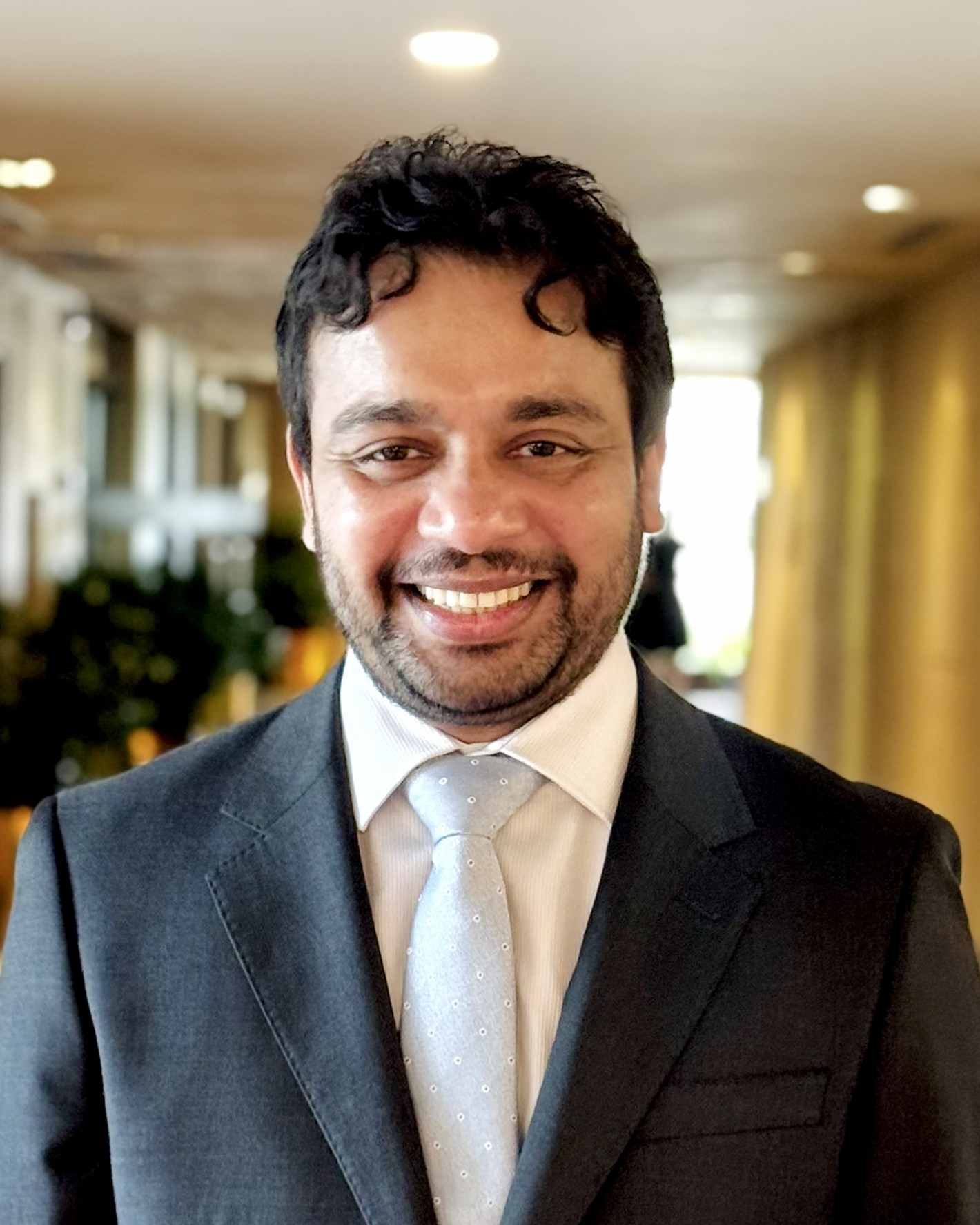}}]{ROSHAN GODALIYADDA}
is a Professor of Electrical and Electronic Engineering at the University of Peradeniya, holding a PhD from the National University of Singapore. His research covers signal and image processing, computer vision, artificial intelligence, and smart grid applications. He has contributed to advanced sensing and data analytics relevant to energy systems, including smart grid integration and monitoring, and has published extensively in leading IEEE journals. He has led and co-led research grants exceeding USD 2 million from international agencies, and his work has received multiple national and university research awards.
\end{IEEEbiography}

\begin{IEEEbiography}[{\includegraphics[width=1in,height=1.25in,clip,keepaspectratio]{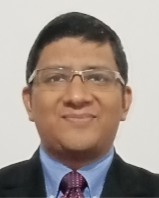}}]{PARAKRAMA EKANAYAKE}
 is a Professor of Electrical and Electronic Engineering at the University of Peradeniya, Sri Lanka, and a Senior Member of IEEE. He holds a PhD in Applied Mathematics from Texas Tech University and a First Class BSc in Engineering from the University of Peradeniya. His research spans artificial intelligence and machine learning, computer vision, hyperspectral and multispectral imaging, smart grids, and data-driven modeling of complex systems. He has led and contributed to nationally and internationally funded multidisciplinary projects and has published extensively in high-impact journals.
\end{IEEEbiography}

\begin{IEEEbiography}[{\includegraphics[width=1in,height=1.25in,clip,keepaspectratio]{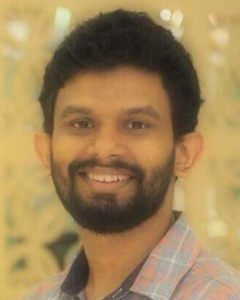}}]{ISURU NAWINNE}
received the B.Sc. degree in engineering from the University of Peradeniya, Sri Lanka, in 2011, and the Ph.D. degree in computer science and engineering from the University of New South Wales, Sydney, Australia, in 2016.
He is currently a Senior Lecturer in Computer Engineering at the University of Peradeniya. He has a keen interest in improving the quality of engineering education and is involved in curriculum development and devising learner-centered delivery methods. His research interests include computer architecture, embedded systems, biomedical engineering, and automation.

\end{IEEEbiography}

\begin{IEEEbiography}[{\includegraphics[width=1in,height=1.25in,clip,keepaspectratio]{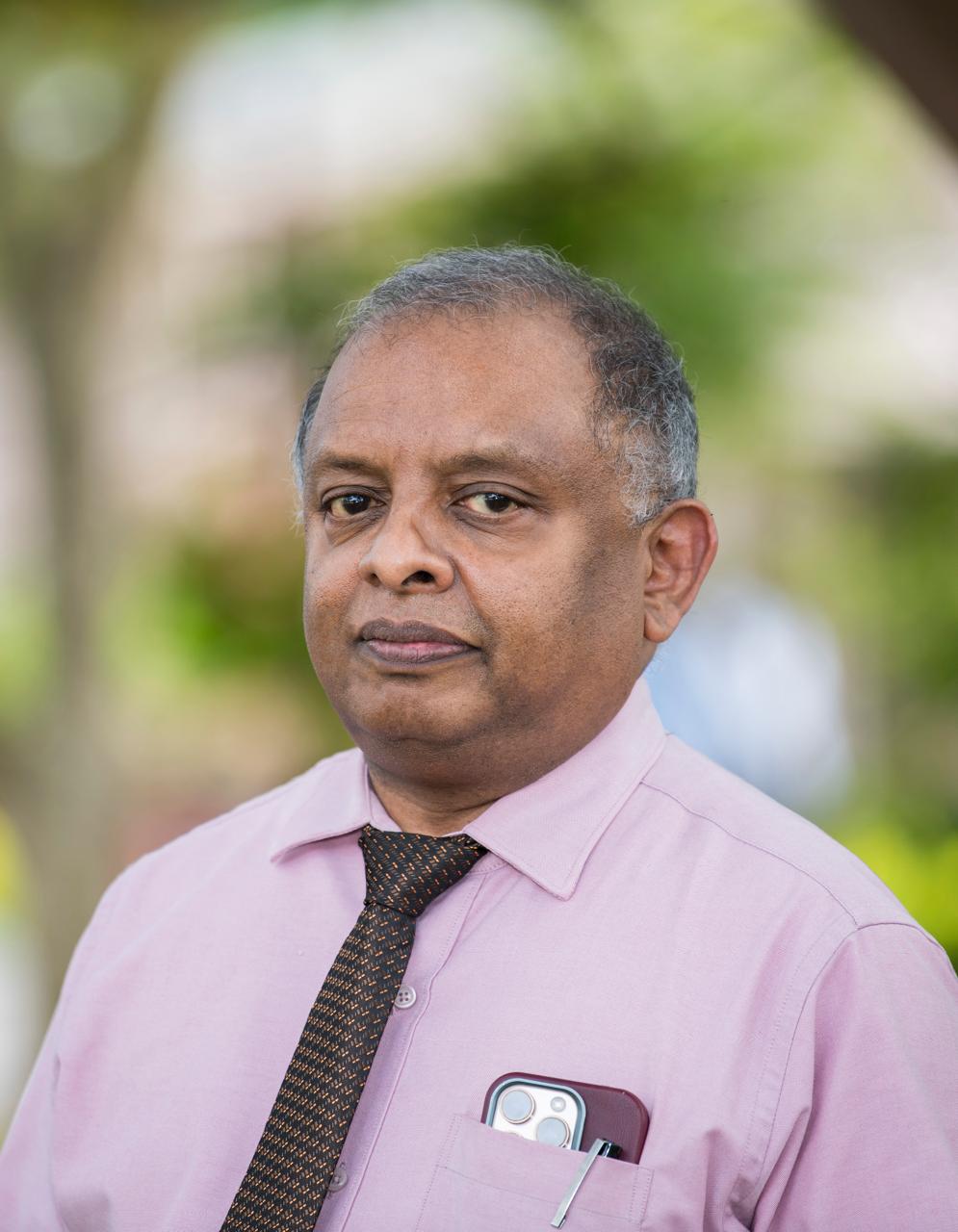}}]{CHATHURA RATHNAYAKE} 
Qualified in Medicine with an MBBS Degree from Faculty of Medicine university of Colombo in 1996. He obtained MS in obstetrics and gynaecology with board certification from Post graduate institute of Medicine university of Colombo in 2004 and Completed MRCOG qualification in UK in 2005. He has been working as a consultant obstetrician and gynaecologist since 2005 and entered department of obstetrics \& gynaecology as a senior lecturer in 2005. He currently holds the chair professorship of the department of obstetrics and gynaecology of the faculty of medicine university of Peradeniya. His main research interests are subfertility, High risk obstetrics and reproductive toxicology and subfertility.
\end{IEEEbiography}

\EOD
\end{document}